# Deadline-aware Power Management in Data Centers[†]

Cengis Hasan[a] and Zygmunt J. Haas[b]
[a]*The University of Edinburgh, School of Informatics*
[b]*Cornell University, School of Electrical and Computer Engineering*

**Abstract**

We study the dynamic power optimization problem in data centers. We formulate and solve the following *offline problem*: in which slot which server has to be assigned to which job; and in which slot which server has to be switched ON or OFF so that the total power is optimal for some time horizon. We show that the offline problem is a new version of generalized assignment problem including new constraints issuing from deadline characteristics of jobs and difference of activation energy of servers. We propose an online algorithm that solves the problem heuristically and compare it with randomized routing.

# 1. Introduction

Energy consumption is one of the most important practical and timely problem associated with data centers for cloud computing. The urgency of this problem has been exposed by both, the governmental agencies and the industry. While policies that consider the physical design of the data centers have been studied in the technical literature rather thoroughly, the operational characteristics of the data centers (e.g., the required performance of the executed applications, such as the type of services and the applications being supported, their QoS requirements, associated background server maintenance processes, etc.) have rarely been ac-counted for. In particular, there is a need to investigate the implications of exploiting the operational characteristics of the data centers in the context of power management policies, as those could lead to potentially significant energy savings and operational cost reduction of a computer cloud.

[†] This work was supported by NSF Grant EAGER: Job-Centered Power Management Policies for Data Centers. The first author was a postdoctoral research associate at Cornell University when this work was carried out.
*Email addresses*: chasan@inf.ed.ac.uk(Cengis Hasan) haas@ece.cornell.edu(Zygmunt J. Haas)

In this work, we study the effect of the deadline characteristics of the jobs (applications) and their requirements on the design of the data centers' power management policies. A particular interest of our study is to solve the dynamic power optimization problem. Assuming a time interval, we seek to minimize the total energy consumed during this time interval. We solve the *offline problem* in which we know the arrival times of jobs and their service demands and deadlines. Actually, the offline problem is the following one:

- consider a discrete time horizon: $T = \{1, \ldots, t_{max}\}$ where an element in $T$ shows a slot,

- a server can be only in ON or OFF state,

- a server can serve only one or none job during a slot,

- a job might be served by only one or none server during a slot,

- a job has to be served within its deadline constraint, i.e. the job's service demand has to be handled within its deadline,

- a server needs $n_{ON}$ consecutive slots (i.e. the setup time) and $E_{ON}$ energy per slot in order to be ON when it is in OFF state,

- a server consumes $E_{slot}$ energy per slot when it serves a job.

Based on these information, we calculate the minimal total energy in the following way:

> *find*: in which slot which server has to be assigned to which job; and in which slot which server has to be switched ON or OFF

Thus, the offline solution becomes a reference for evaluating an online algorithm solving the corresponding problem. Note that an online algorithm can find a solution for only present time. We propose an online algorithm which aims to minimize the total cost per slot. The cost is composed of the energy cost per slot and a weighted sum of deadline and service demand of the job. We calculate every possible assignment of servers to jobs present in the system based on the above cost definition. Then, we need to find which assignment of servers to jobs minimizes the total cost per slot. We show that this problem is exactly the assignment problem which is solvable in polynomial time using the well-known Hungarian algorithm. In case of a server is not assigned to a job, we keep this server ON during a portion of time. This is a similar approach studied in [1].

**1.1. Related Work**

High-level approaches aiming to reduce data center power consumption could be classified in four way [1]: *power-proportionality* which is to find ways in order to guarantee that servers consuming power in proportion to their utilization [2], [3], [4]; *energy-efficient server design* which is to establish the proper server architecture for a given workload [5], [6], [7]; *dynamic server provisioning* which is keeping servers on and off at the right times [8], [9]; *consolidation and virtualization* which is reducing power consumption by resource sharing [10], [11], [12].

The dynamic version of the generalized assignment problem is directly related to the offline problem that we study in this paper. Actually, our problem could be considered as a version of the dynamic generalized assignment problem which includes new additional constraints. In [13], the authors firstly put forward the dynamic generalized assignment problem, and they formulate the continuous-time optimal control model in order to develop an efficient time-decomposition procedure for solving the problem. The discrete time version is studied in [14] by associating a starting time and a finishing time with each task. The authors also use column generation algorithm to compute lower bounds.

## 2. The Offline Problem

**2.1. Problem Description**

Let us represent by $N = \{\emptyset, 1, \ldots, |N|\}$ *the servers*, $J = \{\emptyset, 1, \ldots, |J|\}$ *the jobs* served starting from their arrival slot $t_j$ up to the time characterized by the deadline $t_j + \Delta_j$ where $\Delta_j$ denotes *the deadline* of job $j$. We denote by $T_j = \{t_j, \ldots, t_j + \Delta_j\}$ the maximal lifetime of a job within the system. Consider $t_{max} = \max_{j \in J} t_j + \Delta_j$, then the total time horizon is $T = \{1, \ldots, t_{max}\}$. We would like to calculate *the minimal total energy consumption starting from slot* 1 *until* $t_{max}$.

The duration of a slot is $\tau$. We assume that the deadline of a job can be given by $\Delta_j = n_j \tau$ where $n_j$ is an integer. *The service demand* of any job $j$ is denoted by $w_j$ in terms of processor cycles. Each server is characterized by its *speed* $s_i = \left[\frac{\text{processor cycle}}{\text{second}}\right], \forall i \in N$. So, the total demand that a server can handle during a slot is calculated by $s_i \tau$. If the initial service demand of job $j$ is $w_j$ and server $i$ is assigned to job $j$, then at the end of the slot the service demand of job $j$ reduces to $w_j - s_i \tau$. We assume that a server needs $n_{\text{ON}}$ slots in order to switch ON when it is switched OFF. A server consumes $E_{\text{slot}}$ energy per a slot. If the server is in OFF state and is activated to ON then $E_{\text{ON}}$ energy per slot is needed for it.

Let us define the following binary integer variable,

$$x_{ijt} = \begin{cases} 1, & \text{if server } i \text{ is assigned to job } j \text{ in slot } t \\ 0, & \text{otherwise.} \end{cases} \quad (1)$$

Note that if $x_{\emptyset jt} = 1$, it means that no server is assigned to job $j$ in slot $t$. On the other hand, we also take into consideration that a server is also possible not to be assigned to any job in a slot but it is still ON in that slot, i.e. $x_{i\emptyset t} = 1$. Based on this definition, we assume that any server $i$ in slot $t$ is in OFF state if the following is true:

$$\sum_{j \in J} x_{ijt} = 0, \quad \forall i \in N, \forall t \in T. \quad (2)$$

Besides, we define a new variable $y_{it}$ which has the following meaning:

$$y_{it} = \begin{cases} 1, & \text{if server } i \text{ is OFF and switched ON in slot } t \\ 0, & \text{otherwise.} \end{cases} \quad (3)$$

Note that if $y_{it} = 1$ then, it is always $x_{i\emptyset t} = 1$, otherwise it is not true, i.e. if $y_{it} = 0$ then, $x_{i\emptyset t} = 0$ or 1. This constraint can be given by

$$y_{it} \leq x_{i\emptyset t}, \quad \forall i \in N, \forall t \in T. \quad (4)$$

Thus, the minimal total energy can be calculated by

$$\left\{ \min_{x,y} \left( 0 \sum_{j \in J} \sum_{t \in T} x_{\emptyset jt} + E_{\text{slot}} \sum_{i \in N \setminus \emptyset} \sum_{j \in J} \sum_{t \in T} x_{ijt} \right. \right.$$
$$\left. \left. + (E_{\text{ON}} - E_{\text{slot}}) \sum_{i \in N \setminus \emptyset} \sum_{t \in T} y_{it} \right) \middle| \text{Constraints} \right\}. \quad (5)$$

As we mentioned above, if a server is in OFF state in slot $t-1$, and it is switched ON in slot $t$, then it will spend $n_{ON}$ consecutive slots to setup. Using the above definitions, in order to satisfy that all jobs are fully served, we need the following constraints:

$$\sum_{i \in N} \sum_{t \in T_j} s_i \tau x_{ijt} \geq w_j, \quad \forall j \in J \backslash \emptyset \tag{6}$$

$$\sum_{i \in N} \sum_{t \in T_j} x_{ijt} \leq \frac{\Delta_j}{\tau}, \quad \forall j \in J \backslash \emptyset \tag{7}$$

which means that until the $\frac{\Delta_j}{\tau}$ slot, job $j$ has to be handled. Note that $s_\emptyset = 0$ meaning that if a job is not assigned to a particular server in a slot, its service demand will not be processed during that slot. The constraint allowing at most one job can be assigned to a server in a slot can be given by

$$\sum_{j \in J} x_{ijt} \leq 1, \quad \forall i \in N, \forall t \in T_j. \tag{8}$$

We also need to guarantee that only one or none server can be assigned to a job during a slot (i.e. it is not possible to assign more than one server to a job):

$$\sum_{i \in N} x_{ijt} = 1, \quad \forall j \in J, \forall t \in T_j. \tag{9}$$

We also set $x_{\emptyset\emptyset t} = 0, \forall t \in T$. Moreover, whenever an OFF server is switched ON in slot $t$, then the server cannot be assigned to any job during consecutive $n_{ON}$ slots. If a server is OFF and we would like to switch it ON in slot $t$, then the following constraints must hold:

$$\begin{cases} \text{if } \sum_{j' \in J} x_{ij'(t-1)} = 0 \text{ and } x_{i\emptyset t} = 1, & \sum_{t'=t}^{t+n_{ON}-1} y_{it'} \geq n_{ON} \\ \text{otherwise,} & \sum_{t'=t}^{t+n_{ON}-1} y_{it'} \geq 0. \end{cases} \tag{10}$$

Actually, this if-then constraint can be transferred into the following:

$$\sum_{t'=t}^{t+n_{ON}-1} y_{it'} \geq n_{ON} x_{i\emptyset t} \left(1 - \sum_{j' \in J} x_{ij'(t-1)}\right), \forall i \in N \backslash \emptyset, \forall t \in T. \tag{11}$$

We denote as $z_{it} = x_{i\emptyset t}\left(1 - \sum_{j' \in J} x_{ij'(t-1)}\right)$ the variable which takes value 1 if server $i$ is switched OFF in slot $t-1$ and it is switched ON in slot $t$, otherwise the variable is equal to 0. We are able to transform the multiplication in variable $z_{it}$ as following:

$$\begin{aligned}
z_{it} &\leq x_{i\emptyset t}, & \forall i \in N\backslash\emptyset, \forall t \in T \\
z_{it} &\leq 1 - \sum_{j' \in J} x_{ij'(t-1)}, & \forall i \in N\backslash\emptyset, \forall t \in T \\
z_{it} &\geq x_{i\emptyset t} - \sum_{j' \in J} x_{ij'(t-1)}, & \forall i \in N\backslash\emptyset, \forall t \in T.
\end{aligned} \quad (12)$$

Note that the following must hold always

$$n_{\text{ON}} \sum_{t \in T} z_{it} = \sum_{t \in T} y_{it}, \quad \forall i \in N\backslash\emptyset \quad (13)$$

$$z_{it} \leq y_{it}, \quad \forall i \in N\backslash\emptyset, \forall t \in T. \quad (14)$$

If a server is OFF, it is forbidden to switch it ON last $n_{\text{ON}}$ consecutive slots of last arriving job:

$$\left\{ \text{if } \sum_{j' \in J} x_{ij'(t_{max}-n_{\text{ON}}-1)} = 0, \quad \sum_{t'=t_{max}-n_{\text{ON}}}^{t_{max}-1} y_{it'} \leq 0 \right. \quad (15)$$

which can be transformed to the following form:

$$\sum_{t'=t_{max}-n_{\text{ON}}}^{t_{max}-1} y_{it'} \leq n_{\text{ON}} \sum_{j' \in J} x_{ij'(t_{max}-n_{\text{ON}}-1)}, \forall i \in N\backslash\emptyset. \quad (16)$$

On the other hand, no job can be assigned to an OFF server. The constraint can be given by

$$\text{if } \sum_{j' \in J} x_{ij'(t-1)} = 0, \quad \sum_{j' \in J\backslash\emptyset} x_{ij't} \leq 0 \quad (17)$$

which can be simplified to

$$\sum_{j' \in J\backslash\emptyset} x_{ij't} \leq \sum_{j' \in J} x_{ij'(t-1)}. \quad (18)$$

### 2.2. The Binary Integer Program

The binary integer program based on the above definitions can be collected as following:

$$\min_{x,y} \left( E_{\text{slot}} \sum_{i \in N \setminus \emptyset} \sum_{j \in J} \sum_{t \in T} x_{ijt} + (E_{\text{ON}} - E_{\text{slot}}) \sum_{i \in N \setminus \emptyset} \sum_{t \in T} y_{it} \right) \text{ subject to}$$

$$\sum_{i \in N} \sum_{t \in T_j} x_{ijt} \leq \frac{\Delta_j}{\tau}, \quad \forall j \in J \setminus \emptyset, \tag{19}$$

$$\sum_{i \in N} \sum_{t \in T_j} s_i \tau x_{ijt} \geq w_j, \quad \forall j \in J \setminus \emptyset, \tag{20}$$

$$\sum_{j \in J} x_{ijt} \leq 1, \quad \forall i \in N \setminus \emptyset, \forall t \in T_j, \tag{21}$$

$$\sum_{j' \in J \setminus \emptyset} x_{ij't} \leq \sum_{j' \in J} x_{ij'(t-1)}, \quad \forall i \in N \setminus \emptyset, \forall t \in T, \tag{22}$$

$$y_{it} \leq x_{i\emptyset t}, \quad \forall i \in N \setminus \emptyset, \forall t \in T, \tag{23}$$

$$\sum_{t'=t}^{t+n_{\text{ON}}-1} y_{it'} \geq n_{\text{ON}} z_{it}, \quad \forall i \in N \setminus \emptyset, \forall t \in T, \tag{24}$$

$$z_{it} \leq y_{it}, \quad \forall i \in N \setminus \emptyset, \forall t \in T, \tag{25}$$

$$z_{it} \leq x_{i\emptyset t}, \quad \forall i \in N \setminus \emptyset, \forall t \in T, \tag{26}$$

$$z_{it} \leq 1 - \sum_{j' \in J} x_{ij'(t-1)}, \quad \forall i \in N \setminus \emptyset, \forall t \in T, \tag{27}$$

$$z_{it} \geq x_{i\emptyset t} - \sum_{j' \in J} x_{ij'(t-1)}, \quad \forall i \in N \setminus \emptyset, \forall t \in T, \tag{28}$$

$$\sum_{t'=t_{max}-n_{\text{ON}}}^{t_{max}-1} y_{it'} \leq n_{\text{ON}} \sum_{j' \in J} x_{ij'(t_{max}-n_{\text{ON}}-1)}, \forall i \in N \setminus \emptyset, \tag{29}$$

$$\sum_{i \in N} x_{ijt} = 1, \quad \forall j \in J \setminus \emptyset, \forall t \in T_j, \tag{30}$$

$$n_{\text{ON}} \sum_{t \in T} z_{it} = \sum_{t \in T} y_{it}, \quad \forall i \in N \setminus \emptyset. \tag{31}$$

### 2.3. Linear Programming Relaxation

The optimization that we formulated is a binary integer problem. Since its constraint matrix is not totally unimodular, it is NP-hard to calculate the optimal value in such a case. *The linear programming relaxation* of a binary integer program is the problem that arises by replacing the constraint that each variable must be 0 or 1 by a weaker constraint, that each variable belongs to the interval [0, 1]. That is, for each constraint of the form $x \in \{0,1\}$ of the original integer program, one instead uses a pair of linear constraints $0 \leq x \leq 1$. The resulting relaxation is a linear program.

## 3. Online Heuristics

### 3.1. Assumptions and Cost Function

We consider that the algorithm needs only the information of the current slot $t$ and the previous slot $t-1$. We denote by $\mathcal{J}(t)$ the jobs in the system in slot $t$, the available servers $\mathcal{S}_{\text{ON}}(t)$, the unavailable servers $\mathcal{S}_{\text{OFF}}(t)$, the servers being in the activation process $\mathcal{S}_{\text{A}}(t)$ in slot $t$. Let us also define the *jobs-to-server ratio* per slot as following:

$$\frac{|\mathcal{J}(t)|}{|\mathcal{S}_{\text{ON}}(t)| + |\mathcal{S}_{\text{A}}(t)|}. \tag{32}$$

Moreover, $n_{\text{JA}}$ is the number of jobs to be accumulated in order that we can activate an OFF server.

We assume that

1. if server $i$ is not assigned to a job in slot $t$, then the algorithm starts the waiting time $\tau_{\text{W}}(i, t)$ for server $i$;

2. if $\tau_{\text{W}}(i, t) \geq t_{\text{wait}}$, then the algorithm switches server $i$ from ON to OFF where $t_{\text{wait}}$ is the maximal *wait time*;

3. if $\frac{|\mathcal{J}(t)|}{|\mathcal{S}_{\text{ON}}(t)| + |\mathcal{S}_{\text{A}}(t)|} \geq n_{\text{JA}}$, then activate an OFF server;

Let $\delta_j(t)$ represent *total delay* of job $j$ until slot $t$ since its first occurrence in the system. Let us consider the following cost of assigning server $i$ to job $j$ in slot $t$:

$$c_{ijt} = E_{\text{slot}} + \exp\left(\frac{\delta_j(t-1) - \Delta_j}{\tau}\right) \left(w_j(t-1) - s_i \tau\right)^+ \tag{33}$$

Figure 1. $\exp\left(\frac{\delta_j(t-1)-\Delta_j}{\tau}\right)$ increases monotonically with respect to the time for different deadlines. The $y$-axis is logarithmic.

where by $\frac{\delta_j(t-1)-\Delta_j}{\tau}$, we normalize the difference by slot length $\tau$ letting to characterize the cost of delay. In Figure 1, we depict the change of $\exp\left(\frac{\delta_j(t-1)-\Delta_j}{\tau}\right)$ with respect to the time.

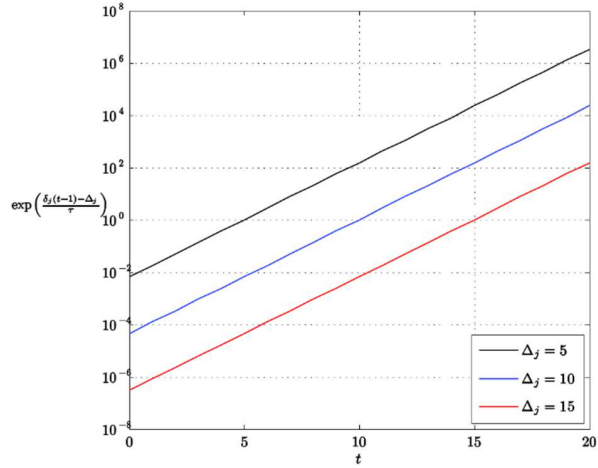

On the other hand, $\left(w_j(t-1) - s_i\tau\right)^+$ is a monotonically decreasing function with respect to the increasing values of slots. The cost of not assigning a server to job $j$ in slot $t$ can thus be given by

$$c_{\varnothing jt} = \exp\left(\frac{\delta_j(t-1)-\Delta_j}{\tau}\right) w_j(t-1). \tag{34}$$

If no job is assigned to server $i$ in slot $t$, the cost is only the energy consumed during a slot

$$c_{i\varnothing t} = E_{\text{slot}}. \tag{35}$$

We set $c_{\varnothing\varnothing t} = 0$ if no server is assigned to no job.

### 3.2. Total Cost Minimization per Slot

Consider that we would like to calculate the total minimal cost per slot based on the above cost definitions. We below show that this problem is exactly the *classical assignment problem* being solvable in polynomial time. Denote as the number of ON servers and the jobs in slot $t$ as $|\mathcal{S}_{\text{ON}}(t)|$ and $|\mathcal{J}(t)|$, respectively. Recall that a server may not be assigned to a job as well as no server can be assigned to a job. Consider some "*virtual servers*" of which number is equal to $|\mathcal{J}(t)|$ in addition to ON servers in slot $t$. We represent by $\varnothing_i$, the $i$th virtual server. Besides, consider some $|\mathcal{S}_{\text{ON}}(t)|$ "*virtual jobs*" in addition to the jobs in slot $t$. We show by $\varnothing^j$, the $j$th virtual job. We define those virtual elements that will substitute the empty set in formulations. Thus, we have totally $|\mathcal{S}_{ON}(t)| + |\mathcal{J}(t)|$ both ON servers and jobs in slot $t$. Below, an example is given.

### 3.3. An Explanatory Example

Assume there are 2 servers and 4 jobs in slot $t$. The cost matrix can be represented as following:

|  |  | \multicolumn{6}{c}{Jobs} |
| --- | --- | --- | --- | --- | --- | --- | --- |
|  |  | 1 | 2 | 3 | 4 | $\varnothing^1$ | $\varnothing^2$ |
| Servers | 1 | $c_{11t}$ | $c_{12t}$ | $c_{13t}$ | $c_{14t}$ | $c_{1\varnothing^1 t}$ | $c_{1\varnothing^2 t}$ |
|  | 2 | $c_{21t}$ | $c_{22t}$ | $c_{23t}$ | $c_{24t}$ | $c_{2\varnothing^1 t}$ | $c_{2\varnothing^2 t}$ |
|  | $\varnothing_1$ | $c_{\varnothing_1 1t}$ | $c_{\varnothing_1 2t}$ | $c_{\varnothing_1 3t}$ | $c_{\varnothing_1 4t}$ | $c_{\varnothing_1 \varnothing^1 t}$ | $c_{\varnothing_1 \varnothing^2 t}$ |
|  | $\varnothing_2$ | $c_{\varnothing_2 1t}$ | $c_{\varnothing_2 2t}$ | $c_{\varnothing_2 3t}$ | $c_{\varnothing_2 4t}$ | $c_{\varnothing_2 \varnothing^1 t}$ | $c_{\varnothing_2 \varnothing^2 t}$ |
|  | $\varnothing_3$ | $c_{\varnothing_3 1t}$ | $c_{\varnothing_3 2t}$ | $c_{\varnothing_3 3t}$ | $c_{\varnothing_3 4t}$ | $c_{\varnothing_3 \varnothing^1 t}$ | $c_{\varnothing_3 \varnothing^2 t}$ |
|  | $\varnothing_4$ | $c_{\varnothing_4 1t}$ | $c_{\varnothing_4 2t}$ | $c_{\varnothing_4 3t}$ | $c_{\varnothing_4 4t}$ | $c_{\varnothing_4 \varnothing^1 t}$ | $c_{\varnothing_4 \varnothing^2 t}$ |

Note that $c_{i\varnothing^j t} = c_{i\varnothing t}$, $\forall j \in \{1,2,\dots,|\mathcal{S}_{ON}(t)|\}$ as well as $c_{\varnothing_i j t} = c_{\varnothing j t}$, $\forall i \in \{1,2,\dots,|\mathcal{J}(t)|\}$. It is obvious that $c_{\varnothing_i \varnothing^j t} = c_{\varnothing \varnothing t} = 0$, $\forall i \in \{1,2,\dots,|\mathcal{J}(t)|\}$ and $\forall j \in \{1,2,\dots,|\mathcal{S}_{ON}(t)|\}$. The total minimal cost in slot $t$ can be calculated by following binary integer program:

$$\min_{x} \sum_{\substack{i \in \mathcal{S}_{ON}(t) \cup \\ \{\varnothing_1,\dots,\varnothing_{|\mathcal{J}(t)|}\}}} \sum_{\substack{j \in \mathcal{J}(t) \cup \\ \{\varnothing^1,\dots,\varnothing^{|\mathcal{S}_{ON}(t)|}\}}} c_{ijt} x_{ijt} \text{ subject to}$$

$$\sum_{\substack{j \in \mathcal{J}(t) \cup \\ \{\varnothing^1,\dots,\varnothing^{|\mathcal{S}_{ON}(t)|}\}}} c_{ijt} x_{ijt} = 1, \quad \forall i \in \mathcal{S}_{ON}(t) \cup \{\varnothing_1,\dots,\varnothing_{|\mathcal{J}(t)|}\},$$

$$\sum_{\substack{i \in \mathcal{S}_{ON}(t) \cup \\ \{\varnothing_1,\dots,\varnothing_{|\mathcal{J}(t)|}\}}} c_{ijt} x_{ijt} = 1, \quad \forall j \in \mathcal{J}(t) \cup \{\varnothing^1,\dots,\varnothing^{|\mathcal{S}_{ON}(t)|}\},$$

$$x_{ijt} \in \{0,1\}, \quad \begin{array}{l} \forall i \in \mathcal{S}_{ON}(t) \cup \{\varnothing_1,\dots,\varnothing_{|\mathcal{J}(t)|}\}, \\ \forall j \in \mathcal{J}(t) \cup \{\varnothing^1,\dots,\varnothing^{|\mathcal{S}_{ON}(t)|}\}. \end{array} \quad (36)$$

The formulation corresponds exactly to the classical assignment problem. The first constraint requires that every server is assigned to exactly one job, and the second constraint requires that every job is assigned to exactly one server. It is known that there is *always an optimal solution* taking integer values even $x_{ijt} \geq 0$. This is because the constraint matrix is totally unimodular. The optimal solution of any assignment problem can be found using well known Hungarian algorithm. Therefore, we utilize this algorithm in calculation of optimal assignments in each slot.

## 4. Online Algorithm

Let us introduce the variables and sets used in the algorithm:

- $\mathcal{J}(t)$: set of jobs not served fully in slot $t$
- $\mathcal{S}_{ON}(t)$: set of available servers in slot $t$
- $\mathcal{S}_{OFF}(t)$: set of unavailable servers in slot $t$
- $\mathcal{S}_{A}(t)$: set of servers in activation process in slot $t$
- $\tau_A(i,t)$: time spent until slot $t$ since activation of server $i$
- $\tau_W(i,t)$: waiting time until slot $t$ of server $i$
- $\tilde{\tau}_A(t)$: vector of time spent until slot $t$ since activation
- $\tilde{\tau}_W(t)$: vector waiting time until slot $t$
- $b_W(i,t)$: the flag bit showing if server $i$ should be in waiting state in slot $t$
- $\tilde{b}_W(t)$: vector of the flag bits
- $\delta_j(t)$: total delay of job $j$ until slot $t$ since its first occurrence in the system
- $w_j(t)$: remained demand of job $j$ in slot $t$
- $A(t)$: assignment matrix in slot $t$

We show by $f(\cdot \,|\, step\ k)$ a set or variable evolved within the algorithm in step $k$. The algorithm works step by step as following:

---

**Algorithm 1**: *Online Algorithm*

1. Input arguments:
$$\left\{ \begin{matrix} \mathcal{J}(t), \mathcal{J}(t-1), \mathcal{S}_{ON}(t-1), \mathcal{S}_{OFF}(t-1), \mathcal{S}_A(t-1), \\ \tilde{\tau}_A(t-1), \tilde{\tau}_W(t-1), \tilde{b}_W(t-1) \\ \left( \delta_j(t), w_j(t), \delta_j(t-1), w_j(t-1) \right)_{\forall j \in \mathcal{J}(t)} \end{matrix} \right\}$$

2. Output arguments:
$$\left\{ \begin{matrix} A(t), \mathcal{S}_{ON}(t), \mathcal{S}_{OFF}(t), \mathcal{S}_A(t), \\ \tilde{\tau}_A(t), \tilde{\tau}_W(t), \tilde{b}_W(t) \\ \left( \delta_j(t+1), w_j(t+1) \right)_{\forall j \in \mathcal{J}(t)} \end{matrix} \right\}$$

3. **set** $\mathcal{S}_{ON}(t) = \mathcal{S}_{ON}(t-1)$, $\mathcal{S}_{OFF}(t) = \mathcal{S}_{OFF}(t-1)$,
$\mathcal{S}_A(t) = \mathcal{S}_A(t-1)$, $\tilde{\tau}_A(t) = \tilde{\tau}_A(t-1)$, $\tilde{\tau}_W(t) = \tilde{\tau}_W(t-1)$,
$\tilde{b}_W(t) = \tilde{b}_W(t-1)$

4. **forall** $s \in \mathcal{S}_A(t)$,
  **if** $\tau_A(s,t) \geq t_{setup}$
  a) Add server $s$ to $\mathcal{S}_{ON}(t)$, i.e. $\mathcal{S}_{ON}(t\,|\,step\ 4) = \mathcal{S}_{ON}(t) \cup s$

       b) Remove server $s$ from $\mathcal{S}_A(t)$, i.e. $\mathcal{S}_A(t|step\ 4) = \mathcal{S}_A(t)\backslash s$
       c) Remove $\tau_A(s,t)$ from $\tilde{\tau}_A(t)$, i.e. $\tilde{\tau}_A(t|step\ 4) = \tilde{\tau}_A(t)\backslash\tau_A(s,t)$
       d) **set** $b_W(s,t|step\ 4) = 0$
       e) **set** $\tau_W(s,t|step\ 4) = 0$
     **endif**
   **endfor**

5. **forall** $s \in \mathcal{S}_{ON}(t|step\ 4)$,
   **if** $b_W(s,t|step\ 4) = 1$ and $\tau_W(s,t|step\ 4) \geq t_{wait}$
      a) **set** $\mathcal{S}_{OFF}(t|step\ 5) = \mathcal{S}_{OFF}(t) \cup s$
      b) **set** $\mathcal{S}_{ON}(t|step\ 5) = \mathcal{S}_{ON}(t|step\ 4)\backslash s$
      c) **set** $\tilde{b}_W(t|step\ 5) = \tilde{b}_W(t|step\ 4)\backslash b(s,t|step\ 4)$
      d) **set** $\tilde{\tau}_W(t|step\ 5) = \tilde{\tau}_W(t|step\ 4)\backslash \tau(s,t|step\ 4)$
   **endif**
   **endfor**

6. **if** $\frac{|\mathcal{J}(t)|}{|\mathcal{S}_{ON}(t)|+|\mathcal{S}_A(t)|} \geq n_{JA}$,
   a) **set** $\mathcal{S}_{OFF}(t|step\ 6) = \{\text{Remove } |\mathcal{J}(t)| - |\mathcal{S}_{ON}(t|step\ 5)| - |\mathcal{S}_A(t|step\ 4)|$ servers from $\mathcal{S}_{OFF}(t|step\ 5)\}$
   b) **set** $\mathcal{S}_A(t|step\ 6) = \mathcal{S}_A(t|step\ 4) \cup \{\mathcal{S}_{OFF}(t|step\ 4)\backslash\mathcal{S}_{OFF}(t|step\ 5)\}$
   **forall** $s \in \mathcal{S}_A(t|step\ 6)$
      a) **set** $\tilde{\tau}_A(t|step\ 6) = \tilde{\tau}_A(t|step\ 4) \cup \tau_A(s,t)$
   **endfor**
   c) **set** $\tilde{\tau}_A(t|step\ 6) = \tilde{\tau}_A(t|step\ 6) + \tau$
   **else**
      a) **set** $\tilde{\tau}_A(t|step\ 6) = \tilde{\tau}_A(t|step\ 4) + \tau$
   **endif**

7. **set** cost matrix $C(t)$
8. Calculate optimal assignments for slot $t$: $A(t) = \text{Hungarian}[C(t)]$
9. **set** $\mathcal{S}_{TEMP} = \emptyset$.
   **forall** $s \in \mathcal{S}_A(t|step\ 6)$,
     **if** server $s$ is not assigned to a job in slot $t$,
       a) add server $s$ to $\mathcal{S}_{TEMP}$, i.e. $\mathcal{S}_{TEMP} = \mathcal{S}_{TEMP} \cup s$
     **else**
       **if** $b_W(s,t|step\ 5) = 1$,
         a) **set** $b_W(s,t|step\ 8) = 0$
         b) **set** $\tau_W(s,t|step\ 8) = 0$
       **endif**
     **endif**
   **endfor**

10. Choose randomly a server $i \in \mathcal{S}_{TEMP}$
    **if** $b_W(i,t|step\ 5) = 0$
      a) **set** $b_W(i,t|step\ 9) = 1$
      b) **set** $\tau_W(i,t|step\ 9) = \tau$
    **else**
11.     a) **set** $\tau_W(i,t|step\ 9) = \tau_W(i,t|step\ 5) + \tau$
    **endif**

12. **if** server $i \in \mathcal{S}_A(t|step\ 6)$ is assigned to job $j \in \mathcal{J}(t)$ in slot $t$
    a) $w_j(t+1) = w_j(t) - s_i\tau$
    **else**
    a) $w_j(t+1) = w_j(t)$
13. **forall** $j \in \mathcal{J}(t)$
    a) **set** $\delta_j(t+1) = \delta_j(t) + \tau$
    **endfor**

## 5. Possible Enhancements in Online Algorithm

Surely, in the current version of online algorithm, the decision of optimal assignments is performed only taking into account the available information preceding two subsequent slots; that can be improved by using some statistical inference techniques where the assignment cost's exponential part given in equation (33) will be updated dynamically. On the other hand, jobs-to-servers ratio given in equation (32) can be transformed to total service demand per total server speed in the following way: $\sum_{j \in \mathcal{J}(t)} w_j(t) / \sum_{i \in \mathcal{S}_{ON}(t)} s_i$. So, the activation of a new server could be decided whenever $\sum_{j \in \mathcal{J}(t)} w_j(t) / \sum_{i \in \mathcal{S}_{ON}(t)} s_i \geq \min_{j \in \mathcal{J}(t)} (t_j + \Delta_j - t)$.

Another enhancement may be to define a new criterion of activation or switching OFF a server which takes into consideration periodically the average number of jobs, service demand, and deadline of jobs. For example, the distribution of hourly number of jobs, service demand, and deadline of jobs is found and applied in the criterion.

## 6. Randomized Routing

We introduce *randomized routing algorithm* which is similar to online algorithm but having a random assignment matrix in step 8. Random assignment matrix is obtained by random permutations.

Table 1. Comparison of online total power cost with random, optimal and relaxed optimal total power cost. $|N|=3$, $|J|=8$.

| Example | $[s_i]_{i \in N}$ | $[w_j]_{j \in J}$ | $[t_j]_{j \in J}$ | $[\Delta_j]_{j \in J}$ | Online Total Power Cost | Randomized Routing Total Power Cost | Optimal Total Power Cost | Relaxed Optimal Total Power Cost |
|---|---|---|---|---|---|---|---|---|
| 1 | [4 2 2] | [4 1 2 5 5 5 1 3] | [2 2 3 3 3 5 5 5] | [3 4 2 2 4 4 4 3] | 3500 | 543960 | 2200 | 1300 |
| 2 | [2 2 4] | [1 1 5 3 2 1 4 1] | [2 2 2 3 4 4 5 6] | [2 4 2 2 4 3 2 3] | 2940 | 93380 | 1800 | 900 |
| 3 | [4 3 2] | [3 1 2 1 3 2 3 2] | [2 2 3 4 4 6 6 6] | [3 2 3 3 4 4 2 4] | 3440 | 137780 | 1600 | 850 |
| 4 | [4 4 4] | [4 4 2 3 3 3 5 2] | [2 2 2 2 4 6 6 6] | [2 3 3 2 2 4 2 4] | 137400 | 138880 | 1800 | 1300 |
| 5 | [4 3 4] | [4 3 1 4 3 1 2 4] | [2 2 3 4 4 4 6 6] | [4 3 4 4 2 2 4 4] | 3800 | 273400 | 1600 | 1100 |

Table 2. Comparison of online total power cost with random online total power cost for different values of parameters. Example 4 in Table 1 is studied.

| | $t_{\text{wait}} = 1$ | | | | | | | |
|---|---|---|---|---|---|---|---|---|
| $n_J$ | 1 | 2 | 3 | 4 | 5 | 6 | 7 | 8 |
| **Online Total Power Cost** | 137400 | 92600 | 92600 | 47400 | 47400 | 47400 | 47400 | 47400 |
| **Random Online Total Power Cost** | 138880 | 138160 | 228200 | 92600 | 137800 | 183000 | 47400 | 183000 |
| | $t_{\text{wait}} = 2$ | | | | | | | |
| $n_J$ | 1 | 2 | 3 | 4 | 5 | 6 | 7 | 8 |
| **Online Total Power Cost** | 3600 | 3600 | 3600 | 3600 | 3600 | 3600 | 3600 | 3600 |
| **Random Online Total Power Cost** | 5400 | 4200 | 4200 | 3600 | 4200 | 6600 | 4800 | 4200 |

## 7. Numerical Results

In each numerical result, we assume that $\tau = 1$ second, $E_{\text{slot}} = 200$ Jouls, $E_{\text{ON}} = 160$ Jouls.

### 7.1. Small Size Examples

In Table 1 and Table 2, the speed of servers, service demands of jobs, arrival slots of jobs, deadlines of jobs follows a uniform distribution taking integer values from [1,4], [1,5], [2,6], [1,4] $\times \tau$, respectively. We assume that all servers are ON, initially.

In Table 1, we assume that $t_{\text{wait}} = 1$ second, $n_J = 1$, $n_{\text{ON}} = 250$. The comparison in Table 1 is based on $|N| = 3$ servers and $|J| = 8$ jobs. Note that even in case of a small size example, the randomized routing algorithm is far from optimal cost. In Example 4, an extreme case occurs where none server is ON when 4th job arrives to the system in slot 5. This happens because of our choice of $t_{\text{wait}} =$ 1 second which causes to switch OFF all servers. Since $n_{ON} = 250$, the activation of a server increases the total power cost.

In Table 2, for different values of $t_{wait}$ and $n_J$, we compare online total power cost with random online total power cost. Note that when $t_{wait}$ increases, both algorithms do better. Increasing $n_J$ could also improve the performance of online algorithm; but, it is not always guaranteed for randomized routing algorithm.

### 7.2. Critical Values of $n_J$ and $t_{wait}$

In Figure 2 and Figure 3, the following parameters are assumed: $t_{wait} = 2$ seconds, $n_J = 5$, $n_{ON} = 10$, and $|J| = 30$. The speed of servers, service demands of jobs, arrival slots of jobs, deadlines of jobs, initial states of servers follow a uniform distribution taking integer values from [2,4], [10,20], [2,200], [10,20] × $\tau$, [0,1], respectively.

In Figure 2, average total power cost with respect to $n_J$ is depicted for $|N| = 1, 2$ and 3. The figure implies that increasing values of $n_J$ tends to result in a positive effect on average total power cost; but, it also shows that for $|N| = 4$, the curve is not monotonically decreasing with respect to $n_J$.

In Figure 3, average number of jobs handled within deadline with respect to $n_J$ is depicted for $|N| = 1, 2$ and 3. It is inevitable that average number of jobs handled within deadline decreases with increasing values of $n_J$.

In Figure 4, we plot how average total power cost changes with respect to $t_{wait}$. We again assume that $n_{ON} = 250$. Figure 4 implies that the average total power could be very dependent on $t_{wait}$. We see that in the considered default values, for $t_{wait} \geq 2$, the average total power cost decreases dramatically. Changing the number of servers and jobs do not affect the critical value of $t_{wait}$.

### 7.3. Comparison of Online Algorithm with Randomized Routing

In Figure 5 and Figure 6, the following parameters are assumed: $t_{wait} = 2$ seconds, $n_J = 5$, $n_{ON} = 250$. Figure 5 plots the change of average total power cost as well as Figure 6 depicts the change of average number of jobs handled within deadline with respect to number of jobs for $|N| = 6, 7$, and 8. The advantage of online algorithm is inevitable in the figures.

## 8. Future Work

As an immediate extension in the offline problem, we can take into account more than two states of the servers, i.e. instead of only ON and OFF modes, we can also consider IDLE, HIBERNATE modes, etc. We can introduce a new time cost in case of a preemption which corresponds to migration a job from a server to another one. Moreover, another constraint can be geographically separated servers as well as we can define hierarchy among jobs and forbid a particular job to be assigned to a particular server. A possible work could also discuss to do flexibility by some factor in deadline constraints of jobs.

## 9. Conclusions

We studied the dynamic power optimization problem in data centers. The corresponding offline problem has been formulated and solved using binary integer programming. We showed that the offline problem is a new version of dynamic generalized assignment problem including new constraints issuing from deadline characteristics of jobs and difference of activation energy of servers. Furthermore, we proposed an online algorithm that solves the problem heuristically and compared it to randomized routing.

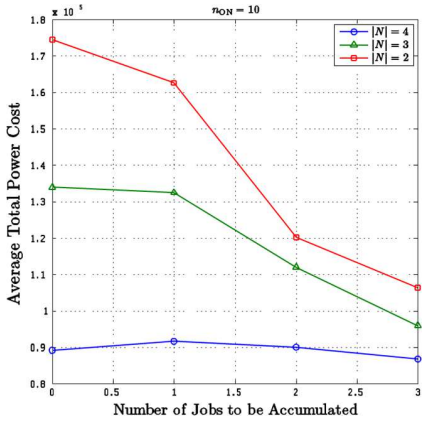
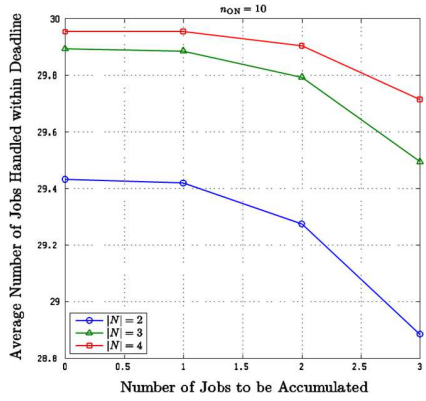
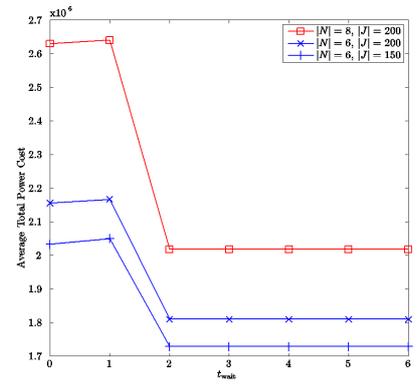

Figure 2. Average total power cost with respect to $n_\text{J}$.

Figure 3. Average number of jobs handled within deadline with respect to $n_\text{J}$.

Figure 4. Average total power cost with respect to $t_\text{wait}$.

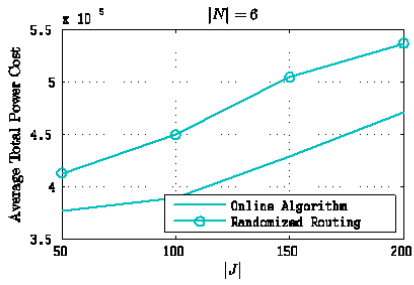
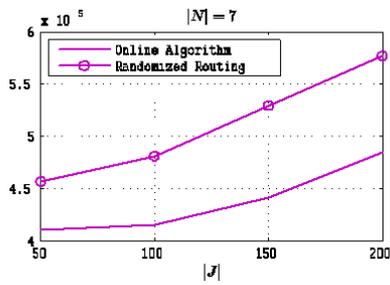
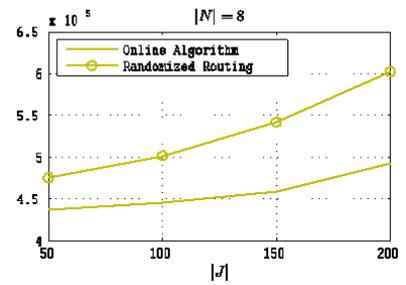

Figure 5. Comparison of online algorithm with randomized routing in terms of average total power cost with respect to $|J|$.

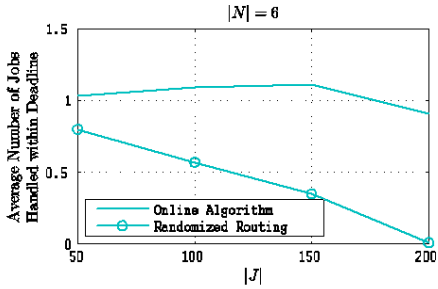
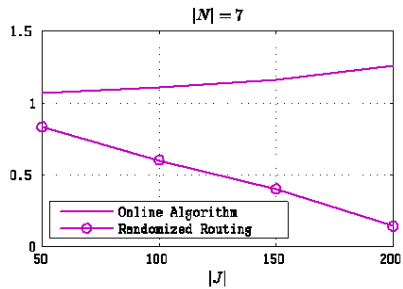
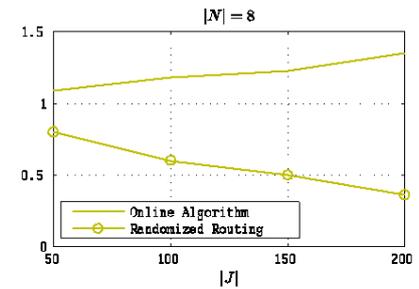

Figure 6. Comparison of online algorithm with randomized routing in terms of average number of jobs handled within deadline with respect to $|J|$.

18